\documentclass[runningheads]{llncs}

\usepackage{float}

\usepackage{pdflscape} %landscape
\usepackage{graphicx} % Required for inserting images
\usepackage{amsmath}
\usepackage{makecell}
\usepackage{amssymb}
\usepackage{hyperref}
\usepackage{authblk}  % Package for author affiliations
\usepackage[textsize=tiny]{todonotes}

\usepackage{bbm}

\date{September 2024}

\usepackage{color, colortbl}
\usepackage{stfloats}
\usepackage[most]{tcolorbox}
\usepackage{url}
\usepackage{xspace}
\usepackage{todonotes}
\usepackage{tabularx}
\usepackage[inline,shortlabels]{enumitem}
\usepackage[shortcuts]{extdash}
\usepackage{booktabs}
\usepackage[english]{babel}
\usepackage{blindtext}
\usepackage{multirow} 
\usepackage{subcaption}
\usepackage{threeparttable}
\usepackage{placeins}
\usepackage{lscape}

\newcommand{\dex}{DEX\xspace}
\newcommand{\dexs}{DEXs\xspace}

\newcommand{\amms}{AMMs\xspace}
\newcommand{\defi}{DeFi\xspace}

\newcommand{\lps}{LPs\xspace}

\begin{document}

\title{What Drives Liquidity on Decentralized Exchanges? Evidence from the Uniswap Protocol}

\titlerunning{What Drives Liquidity?}

%\author{Anonymous Authors}
%\institute{}

\author{
Brian Zhu \inst{1} \and
Dingyue Liu \inst{2} \and
Xin Wan \inst{2} \and
Gordon Liao \inst{3}, \\ 
\vspace{-1em}
Ciamac Moallemi \inst{1} \and
Brad Bachu \inst{2}
}

\institute{
    Columbia University \\
    \email{bzz2101@columbia.edu, ciamac@gsb.columbia.edu} \and 
    Uniswap Labs \\
    \email{kite.liu@uniswap.org, xin@uniswap.org, brad.bachu@uniswap.org} \and 
    Circle Internet Financial \\
    \email{gordon@circle.com}}

\authorrunning{Zhu et al.}

\maketitle

\vspace{-1em}

\begin{abstract}
    We study liquidity on decentralized exchanges (DEXs), identifying factors at the platform, blockchain, token pair, and liquidity pool levels with predictive power for market depth metrics. 
We introduce the \textit{v2 counterfactual spread} metric, a novel criterion which assesses the degree of liquidity concentration in pools using the ``concentrated liquidity'' mechanism, allowing us to decompose the effect of a factor on market depth into two channels: total value locked (TVL) and concentration. 
We further explore how external liquidity from competing DEXs and private inventory on DEX aggregators influence market depth. 
% We further explore how external liquidity on DEX aggregators from fillers with private inventory influence market depth. 
% We find, for moderately-sized swaps, that (i) gas prices, returns, and volatility affect spreads primarily through concentration, (ii) internalization of swaps by private liquidity affects spreads primarily through TVL, and (iii) fee revenue, markout, and DEX competition affect spreads through both channels.
We find that (i) gas prices, returns, and a DEX's share of trading volume affect liquidity through concentration, (ii) internalization of order flow by private market makers affects TVL but not the overall market depth, and (iii) volatility, fee revenue, and markout affect liquidity through both channels.
% These results are consistent with previous work on liquidity provision in DEXs.
% \keywords{Blockchain  \and DeFi \and Automated market makers \and Liquidity}
\end{abstract}
%

% ------------------------------

\section{Introduction}\label{sec:intro}

Liquidity plays a fundamental role in financial markets, serving as a critical determinant of market efficiency and stability. 
This is particularly evident in traditional finance (TradFi), where liquidity impacts execution prices, price discovery, and overall market robustness.
% An important measure of market liquidity in TradFi is the bid-ask spread: the difference between the highest price at which a market maker is willing to buy (bid) and the lowest price at which one is willing to sell (ask), with narrower spreads indicating higher liquidity and more efficient price discovery.
Extensive research has explored how factors such as asset volatility and investor behavior shape liquidity in TradFi.
However, the evolving nature of decentralized finance (\defi) introduces dynamics for liquidity provision that remain under-explored.

% \cite{kyle1985continuous,amihud1986asset,pastor2003liquidity,chordia2001market,hasbrouck2001common,glosten1985bid}

% Decentralized exchanges (\dex) introduce novel paradigms for liquidity provision and trading. 
% Built on blockchain technology, \dexs operate via automated market makers (\amms).
% Unlike order-book-based centralized exchanges (CEXs), \amms rely on liquidity pools and pricing functions. Any individual can be a liquidity provider (LP) to the pool by depositing tokens into the pool, which can be withdrawn at any time.
% Traders then directly interact with the liquidity pool, where trading prices are then determined by the AMM pricing function. 
% This decentralized approach significantly alters the mechanics behind liquidity provision and trading: LPs are not required to provide quotes as they are automatically determined by the pricing function, and trader's orders do not need to be matched as they trade directly with the liquidity pool.

Decentralized exchanges (\dex) introduce novel paradigms for liquidity provision and trading, utilizing \textit{liquidity pools} and \textit{pricing functions} as opposed to limit order books. 
Understanding the dynamics of liquidity in \dexs under this new paradigm is not only important for traders and investors, but also for the design and development of DEXs.
While a substantial body of literature exists in TradFi regarding liquidity, there is a pressing need for more research on the idiosyncratic elements of liquidity provision in \dexs. 
This paper addresses this gap by investigating the forces that drive liquidity and market depth in \dexs, contributing to both academic discourse and practical applications in DeFi.

% Measuring market liquidity: an introductory survey

One recent development in DeFi has been the rise of liquidity aggregators, which combine liquidity from on- and off-chain sources to deliver better execution prices for trades. While research has shown that these services improve prices for traders, their impact on liquidity provision in AMMs is less studied. We answer this question by analyzing if and how these services affect on-chain liquidity.

We focus on liquidity dynamics within the Ethereum ecosystem, examining pools on the Uniswap v3 protocol deployed on the Ethereum Mainnet (L1) and Layer 2 (L2) networks.
As the primary blockchain for decentralized applications, Ethereum is host to a variety of \dexs, with Uniswap standing out as the leading platform in trading volume, total value locked (TVL), and user adoption. While our analysis focuses on Uniswap v3, we show that our framework is applicable to a broader class of AMMs, including those used on Uniswap v2 and v4.
% Uniswap, which operates using Automated Market Maker (\amm) protocols, plays a central role in the \defi ecosystem by facilitating permissionless trading and liquidity provision.
% By studying liquidity on Uniswap V3 pools across the Ethereum mainnet and Layer 2 solutions, we aim to understand how innovation like centralized liquidity, transactions costs, new chain deployment influence liquidity behavior.

\paragraph{\textbf{Our Contributions.}} 
The results of our analysis offer valuable insights into the determinants of liquidity in \amms. 
The key contributions of this paper are:

\begin{enumerate}
\item 
\textbf{Identifying on-chain predictive factors for liquidity:} We identify factors on period $t$ that forecast various market depth metrics on period $t+1$. 
We find that gas prices, token pair returns and volatilities, and in-pool fee revenue and markout have significant explanatory power on future market depth, consistent with prior theoretical results.
\item \textbf{Introducing a novel metric to evaluate liquidity concentration:} By developing the \textit{counterfactual v2 spread} metric, we present a novel technique to assess the concentration of liquidity in Uniswap v3 pools. 
This allows us to identify the channel(s) in which changes to market depth occur, whether through the \textit{deployment} and/or \textit{concentration} of liquidity.
\item  \textbf{Understanding impacts of external liquidity:} We examine the impact of liquidity sources outside Uniswap v3 pools, focusing on competing DEX liquidity and off-chain liquidity used by aggregators. We find that a higher competitor market share negatively impacts liquidity, while more internalization by fillers using private off-chain inventory has no significant impact on overall liquidity.

% increased internalization by fillers with private off-chain inventory negatively impacts on-chain liquidity in Uniswap v3 pools, suggesting that liquidity fragments across DEXs. 
% and dries up in the presence of private market makers.
%We examine the impact of introduction of off-chain liquidity used by services like 1Inch Fusion\cite{1inchNetwork}, CowSwap\cite{cowswap}, or UniswapX\cite{unix}, and show that it negatively impacts on-chain liquidity provision.
% \item
% \textbf{Quasi-Experimental Analysis of Market Events}: We employ quasi-experimental approaches to examine the impact of various market events on liquidity provision. 
% Specifically:
% \begin{itemize}
%     \item Protocol deployments on new chains: These events have a temporary, but not permanent, effect on liquidity in existing chains.
%     \item Localized temporary effects: The effect is more pronounced on the chain where the protocol was most recently deployed.
%     \item Introduction of external liquidity sources: The introduction of outside liquidity (such as off-chain liquidity utilized by services like 1Inch Fusion, CowSwap, or UniswapX) negatively impacts on-chain liquidity provision.
% \end{itemize}
\end{enumerate}

\paragraph{\textbf{Related Literature.}} Our paper contributes to the literature on liquidity provision in DEXs. Some studies focus on incentives for/against liquidity providers (LPs). Lehar and Parlour \cite{lehar2021decentralized} as well as Capponi and Jia \cite{capponi2024liquidity} study equilibrium in liquidity pools, showing that volatility arbitrage risk causes LPs to exit pools. Capponi, Jia, and Zhu \cite{capponi2023paradox} analyze the phenomenon of just-in-time liquidity \cite{xin2023}, showing that it may lead to shallower pools by taking fees away from and leaving toxic order flow to passive LPs. An important factor behind these incentives are the losses incurred by LPs. One popular loss metric is \textit{impermanent loss}, which has been widely discussed \cite{heimbach2022risks,kim2024comparison,li2024implied,li2024yield}. Milionis et al. \cite{milionis2022lvrfees,milionis2022automated} introduce and study \textit{loss-versus-rebalancing}, a forward-looking risk measure that incorporates volatility and market conditions. % Fang \cite{fang2024amm} leverages LP profitability as evidence of liquidity misallocation in AMMs, and shows that this misallocation declines over time.

Other studies focus on liquidity in AMMs that use \textit{concentrated liquidity}, e.g.\ Uniswap v3.
Lehar, Parlour, and Zoican \cite{lehar2024fragmentation} find that large (small) LPs prefer low-fee (high-fee) pools on Uniswap v3 and adjust their positions (in)frequently. 
Lyandres and Zaidelson \cite{lyandres2024efficiency} examine capital allocation on Uniswap v3, finding that market efficiency casually impacts capital efficiency. 
Cartea et al.\ \cite{cartea2024v3} and Fan et al.\ \cite{fan2024v3} study strategies for liquidity provision on Uniswap v3 in terms of various market parameters. 

% Aoyagi et al.\ \cite{aoyagi2024} identifies and analyzes a tradeoff between toxicity and competitiveness in Uniswap v3 pools.

We contribute to this literature by providing the first comprehensive empirical analysis regarding determinants of liquidity and market depth on DEXs. We consider the effects of multiple factors on liquidity simultaneously, with our sample spanning three years and across multiple blockchains.

Closest to our \textit{counterfactual v2 spread (Cv2S)} metric for liquidity concentration is the \textit{capital allocation efficiency (CAE)} metric of \cite{lyandres2024efficiency}. While CAE is dependent on the trades that occurred in the pool during the calibration period, Cv2S is a function of trade size and independent of other trades. Thus, two pools with the same ``liquidity landscape'' can have different CAE values, but always have the same Cv2S given trade size.

Our paper also contributes to the literature on informed trading taking place in DEXs. Capponi, Jia, and Yu \cite{capponi2023price} show that trades with higher priority fees contain more information and have a higher price impact. Klein et al.\ \cite{klein2023price} analyze information contained in both trade and liquidity events on DEXs, finding evidence of heterogeneity in price impact across several dimensions. We contribute by showing that informed trading within a pool, proxied by markout, has a negative effect on market depth.

Another contribution of our paper is to the literature on liquidity in off-chain exchanges for cryptocurrencies, i.e.\ centralized exchanges (CEXs) and DEX aggregators. Brauneis et al.\ \cite{brauneis2021measure} study liquidity on CEXs, finding that returns and volume have predictive power on liquidity. 
% Hu and Zhang \cite{huzhang2024competition} study competition on CEXs, finding that more competition complements volume on existing exchanges.
Bachu, Wan, and Moallemi \cite{bachu2024quantifying} provide empirical evidence of DEX aggregators improving prices for traders. We contribute by finding that more internalization of order flow by aggregators negatively affect on-chain liquidity in pools.
% (such as 1inch Fusion~\cite{1inchNetwork}, CowSwap~\cite{cowswap}, and UniswapX~\cite{unix}) 
\section{Background}

% \subsection{Decentralized Exchanges}
% Decentralized exchanges (\dexs) represent a transformative shift in the way individuals trade assets.
% The key difference between \dexs and centralized exchanges (CEXs) lies in the removal of custody and control over user funds.
% In a \dex, users maintain ownership of their assets through non-custodial wallets, and transactions are facilitated via smart contracts.
%Unlike centralized exchanges (CEXs), where an intermediary manages order books and facilitates transactions, \dexs operate autonomously on blockchain networks, allowing peer-to-peer trading without the need for a trusted third party.
%The main idea behind \dexs is to offer permissionless and trustless trading by utilizing smart contracts and automated market makers (\amms).

% The core mechanism that drives most \dexs is the Automated Market Maker (\amm) model, which replaces traditional order books with liquidity pools. 
% These pools are funded by liquidity providers (\lps), who deposit pairs of tokens into smart contracts. 
% When users trade on a \dex, they interact with these pools rater than a counterpart, ensuring continuous liquidity regardless of trading volume.

\paragraph{\textbf{Automated Market Makers.}} AMMs use liquidity pools and algorithmic pricing functions to facilitate the on-chain exchange of tokens. When LPs deposit tokens into a pool, they receive pool tokens that indicate their stake of the pool and determine the amount they withdraw. Traders typically have to pay a fee proportional to the trade size; this fee is distributed pro-rata among LPs by their stake and incentivizes them to stay in the pool.  

Many AMMs are constant function market makers (CFMMs), which requires post-trade pool reserves to be on the same level set of the pricing function as pre-trade reserves. For example, in a pool with $X$ tokens X and $Y$ tokens Y, a trade of $\Delta_X$ tokens X for $\Delta_Y$ tokens Y must satisfy the relation $F(X+\Delta_X,Y+\Delta_Y)=F(X,Y)$, where $F:\mathbb{R}^2_+\to\mathbb{R}$ is the pricing function. AMM protocols using this design include Uniswap v2, Curve, and Balancer.

The ``concentrated liquidity'' (CL) mechanism for AMMs~\cite{adams2021uniswap}, first pioneered by Uniswap v3,  allows \lps to choose the price range in which their liquidity is active.
This innovation provides higher capital efficiency, but also introduces new complexities for \lps in terms of managing risk and exposure, as out-of-range LP positions do not earn fees and a poorly concentrated pool may increase trading costs. 
Protocols also using CL include PancakeSwap v3 and SushiSwap v3.

\paragraph{\textbf{DEX Aggregators.}}

The success of AMMs has lead to a proliferation in DEXs, with there being over 100 DEXs at the time of writing. 
This growth has led to the fragmentation of liquidity across multiple DEXs. 
In response, protocols such as 1inch Fusion, CowSwap, and UniswapX have introduced new methods to handle order flow, leveraging liquidity from various on-chain sources and off-chain private market makers (PMMs), to optimize trading outcomes for users in a fragmented ecosystem. 

DEX aggregators process order flow from their interfaces and allow specialized users, including PMMs, to determine the ordering and/or routing of trades to achieve better execution prices, most commonly implemented via order flow auctions (OFAs). These OFAs can have varying formats: for example, CowSwap uses batch auctions, whereas 1inch Fusion and UniswapX use Dutch auctions.

% However, these services have important implications for on-chain liquidity provision. 
% From one single DEX's point of view, competing DEXs take away swap volume, and aggregator services reduce the volume of swaps that directly interact with AMMs by routing swaps away to be filled by private liquidity sources.
% These factors potentially influence spreads, overall liquidity, and the revenue generated by on-chain LPs. 

% To better understand the effects that DEX competition and private liquidity have on liquidity provision on-chain, we modify the baseline regression model to include variables that capture the volume of swaps taking place on other DEXs or filled by private liquidity due to aggregator routing.
% We track swap volume (i) on other DEXs and (ii) routed through aggregators, isolating swaps filled completely by private liquidity by comparing transaction hashes with on-chain events.
% From this, we compute the \textit{DEX competition ratio} and \textit{internalization ratio}, defined as follows.

\section{Data}
\label{sec:data}

% \subsection{Sample Construction}

We use publicly available Uniswap v3 data from May 5, 2021 to July 31, 2024. The liquidity pools in our sample, shown in Table 1, are selected as follows:
\begin{itemize}
    \item Obtain the top 4 blockchains by average trading volume through the sample period. For each selected blockchain, obtain the top 100 pools by average trading volume through the sample period.
    \item Select the pools corresponding to the token pair and fee tier combinations appearing in all four top-100 lists.
\end{itemize}

\vspace{-1.5em}

\setlength{\tabcolsep}{0.25em} % for the horizontal padding
\renewcommand{\arraystretch}{1.15}% for the vertical padding

\begin{table}[h]
    \centering
    \begin{tabular}{|c|c|c|c|c|}
        \hline
        \textbf{Pair $\backslash$ Network} & Ethereum (L1) & Arbitrum (L2) & Optimism (L2) & Polygon (L2)  \\
        \hline
        CRV--WETH & 30 bps & 30 bps & 30 bps & 30 bps  \\
        \hline
        DAI--WETH & 30 bps & 30 bps & 30 bps & 30 bps  \\
        \hline
        LDO--WETH & 30 bps & 30 bps & 30 bps & 30 bps  \\
        \hline
        LINK--WETH & 30 bps & 30 bps & 30 bps & 30 bps  \\
        \hline
        USDC--WETH & 5, 30 bps & 5, 30 bps & 5, 30 bps & 5, 30 bps  \\
        \hline
        WBTC--WETH & 5, 30 bps & 5, 30 bps & 5, 30 bps & 5, 30 bps  \\
        \hline
        WETH--USDT & 5, 30 bps & 5, 30 bps & 5, 30 bps & 5, 30 bps  \\
        \hline
    \end{tabular}
    \vspace{1em}
    \caption{Liquidity Pools Included in Sample.}
\end{table}\label{table:pools}

\vspace{-3.5em}

% As of August 2024, these pools contains \textcolor{blue}{\$xx} of trading volume during the analysis period, and it is \textcolor{blue}{xx\%} of the total trading volume on Uniswap. 
% And these pools on average contains \textcolor{blue}{\$xx} of liquidity, \textcolor{blue}

\subsection{Liquidity Metrics}

% For each pool in our sample, we compute effective spreads given a fixed trade size and TVL at a daily frequency.
% Spread data is acquired through Uniswap v3 quoter contracts.\footnote{ \url{https://docs.uniswap.org/contracts/v3/reference/periphery/lens/Quoter}} TVL data is computed by aggregating liquidity and swap events from public blockchain data on Dune Analytics. A more detailed description of the variables follows:

\paragraph{\textbf{Effective Spread.}} We compute the effective spread, which we define as the difference in quoted price between buying and selling a fixed amount of a given token in a liquidity pool, minus transaction fees. 
By using quoted prices rather than execution prices and subtracting out transaction fees, we ignore the impacts of MEV- and fee-related slippage (see \cite{adams2024cost} for example), which isolates the effect of liquidity on the trading costs.
Effective spreads are a key measure of liquidity, representing the difference between buying and selling an asset in a market, with a smaller (larger) effective spread indicates a deeper (shallower) market.
% In our analysis, we compute the effective spread for fixed trade sizes of $\Delta$ WETH for each liquidity pool in the sample as a proxy of the liquidity to assess how liquidity provision evolves over time.
% The spread is a key measure of liquidity, representing the difference between the bid and ask prices in a market. 
% In \dexs like Uniswap, the spread can be interpreted as the price difference between buying and selling a fixed amount of a given token in a liquidity pool. 
% A smaller spread indicates higher liquidity and more efficient price discovery, while a larger spread suggests lower liquidity and higher trading costs. 
% In our analysis, we compute the daily spread for each liquidity pool in the sample as a proxy of the liquidity to assess how liquidity provision evolves over time.

We acquire quoted prices from Uniswap v3 quoter contracts.\footnote{ \url{https://docs.uniswap.org/contracts/v3/reference/periphery/lens/Quoter}} These contracts have functions to obtain quotes for buying or selling a token at historical blocks, allowing the user to specify blockchain, token pair, fee tier, and swap size. 
For a given trade size of $\Delta$ WETH, we compute the relative difference in quoted price between buying and selling $\Delta$ WETH minus twice the fee tier $f$. 
Using Uniswap v3's quoter contract, the \texttt{ExactOutput} function yields the ask price, denoted $A$, and the \texttt{ExactInput} function yields the bid price, denoted $B$.
For each day in our sample period, we obtain quotes every six hours (for four samples per day), and take the spread for the day as the average of these measurements. 

Formally, the normalized effective spread in basis points attributed to market depth at day $t$ on a given pool with fee tier $f$ is
\begin{align*}
    \textsf{v3Spread}_t^{pool} = 10^4\times\left(\frac{1}{4}\sum_{i\in[4]} \frac{A^{pool}_{t,i}-B^{pool}_{t,i}}{\frac{1}{2}(A^{pool}_{t,i}+B^{pool}_{t,i})}-2f\right)
\end{align*}
where $i$ indexes the in-day samples. Since quoted prices contain fees, we subtract $2f$ from the average and multiply by $10^4$ to obtain basis points. For our selected pools, $f$ is 0.0005 or 0.003, corresponding to 5 or 30 basis points, respectively. % In our analysis, we use a trade size of $\Delta=1$ WETH.

\paragraph{\textbf{Total Value Locked.}} Total value locked (TVL), the US dollar value of a pool's token reserves, is another important measure of liquidity in DEXs.
In CFMMs, the TVL is a perfect signal of market depth as execution prices are computed directly based on the pool reserves. 
In AMMs using CL like Uniswap v3, however, TVL is a noisier signal as market depth also depends on how those reserves are concentrated around the pool's current tick. 
Since fee revenue and risk are shared pro-rata on the Uniswap protocol, TVL is also a benchmark that other pool metrics can be normalized against. 

We compute TVL by aggregating mint (deposit), burn (withdrawal), and swap events on liquidity pools, keeping track of how token quantities in each pool vary over time. At the end of each day $t$, we sum the number of each token in a liquidity pool weighted by the end-of-day token price in USD to arrive at the end-of-day TVL.

\vspace{-0.75em}

\begin{table}[H]
    \centering
    \begin{tabular}{ccccc}
\toprule
 & $\log\textsf{\,v3Spread}$ (L1) & $\log\textsf{\,TVL}$ (L1) & $\log\textsf{\,v3Spread}$ (L2) & $\log\textsf{\,TVL}$ (L2) \\
\midrule
$N$ & 11371 & 11371 & 27069 & 27069 \\
Mean & -0.6944 & 17.976 & 2.6869 & 13.7407 \\
S.D. & 2.1329 & 1.9306 & 2.3198 & 1.9985 \\
Skew & 1.7493 & -1.5520 & 0.1135 & -0.1417 \\
25\% & -1.9277 & 17.5133 & 0.9339 & 12.3722 \\
50\% & -1.2828 & 18.3872 & 2.5252 & 13.7500 \\
75\% & -0.1236 & 19.5271 & 4.1776 & 15.2480 \\
\bottomrule
\end{tabular}
\vspace{1em}
    \caption{Summary Statistics for Liquidity Metrics.}
    \vspace{-1em}
\end{table} \label{tab:summstats}

\paragraph{\textbf{Summary Statistics.}} The summary statistics of our liquidity metrics are displayed in Table 2. We highlight the differences between liquidity in pools on L1 (Ethereum) and L2 networks. Effective spreads are lower, less dispersed, and more right-skewed on L1 pools relative to L2 pools. TVL is higher, less dispersed, and more left-skewed on L1 pools than L2 pools. Notably, both liquidity metrics are relatively normally distributed for L2 pools in our sample. Relevant data visualizations for these liquidity metrics are in Appendix \ref{sec:datavisualization}. 

\subsection{Independent Variables}

The set of factors to regress our liquidity metrics is motivated by previous theoretical and empirical research on liquidity in DEXs, from which several factors consistently appear: gas price at the blockchain level, price returns and volatility at the token pair level, and adverse selection (informed trading) and fee revenue (noise trading) at the pool level. These are summarized in Table 3, and generally agree with each other on the direction effects of the variables on liquidity.

\vspace{-1em}

\begin{table}[H]
    \centering
    \begin{tabular}{|c|c|c|}
        \hline
        \textbf{Variable} & \textbf{Prediction/Finding (Setting)}  \\
        \hline
        gas price & \cite{lehar2024fragmentation,li2024yield}: $\nearrow$ gas price $\implies$ $\searrow$ rebalancing frequency (v3) \\
        \hline
        returns & \cite{cartea2024v3}: returns have ambiguous effects on price ranges (v3) \\
        \hline
        volatility & \makecell{\cite{lehar2021decentralized}: $\nearrow$ volatility $\implies$ $\searrow$ pool size (v2) \\ \cite{capponi2024liquidity,cartea2024v3}: $\nearrow$ volatility $\implies$ wider ranges (v3)} \\
        \hline
        fee revenue & \makecell{\cite{capponi2024liquidity,capponi2023paradox,lehar2021decentralized}: $\nearrow$ fee revenue $\implies$ $\nearrow$ pool size (v2) \\ \cite{cartea2024v3}: $\nearrow$ fee revenue $\implies$ narrower ranges (v3)} \\
        \hline
        adverse selection & \makecell{\cite{capponi2024liquidity,capponi2023paradox}: $\nearrow$ adverse selection $\implies$ $\searrow$ pool size (v2) \\ \cite{capponi2024liquidity}: $\nearrow$ adverse selection $\implies$ narrower ranges (v3)} \\
        \hline
    \end{tabular}
    \vspace{1em}
    \caption{Factors Affecting Liquidity Provision Studied by Previous Works.}
\end{table}\label{table:works}

\vspace{-2.5em}

We enhance these studies by empirically testing theoretical predictions and verifying empirical findings in a setup with more variables to rule out confounding. Our counterfactual v2 spread metric also allows us to test implications for both Uniswap v2- and v3-like AMMs with only Uniswap v3 data. We thus select gas prices, returns, volatility, markout, and fee revenue as a baseline set of factors, with markout as a proxy for adverse selection. We compute these variables at a daily frequency, using publicly available data from Dune Analytics. We collect data on gas prices per transaction on each blockchain, token price data from centralized exchanges (CEXs), and data on liquidity and swap events occurring on each pool. More detailed descriptions of each variable follow.

% including timestamps, transaction hashes, and token in/out amounts for each mint, burn, and swap, as well as pool prices.
% Summary statistics and data visualizations are included in Appendix~\ref{appendix:summarystat}:
% A more detailed description of the variables follows:

\paragraph{\textbf{Gas Prices.}} 
We compute the average gas price, in USD, of all transactions on a given chain for the current day $t$. 
% This variable reflects the cost of interacting with the Ethereum Mainnet or Layer 2 networks. 

\paragraph{\textbf{Log-Returns.}} 
Let $\{p_t\}$ be the price ratio of the token pair traded on a pool in our sample, in units of the other token per WETH and measured via CEX prices, at day $t$. Log-returns at time horizon $h_r$ are then given by
\begin{align*}
    \textsf{LogReturns}_t^{pair} = 100\times\log\,(p_t/p_{t-h_r}).
\end{align*}

\paragraph{\textbf{Volatility.}} 
We compute the annualized volatility of the token price ratio during day $t$, using 15-minute intervals to obtain returns.
This choice of time interval captures the fine-grained intra-day price variability while reducing the influence of microstructure noise present in shorter intervals. 
% Volatility is moderately correlated with absolute log-returns and has been proposed as a proxy for LP profitability~\cite{milionis2022automated}; 
% we include both absolute log-returns and volatility to identify how each channel affects liquidity.

\paragraph{\textbf{Fee Revenue.}}
The pro-rated fee revenue (or \textit{pool APR}), computed by dividing total fees accrued to a pool from a day's swaps by the end-of-day pool TVL, is
\begin{align*}
    \textsf{FeeRevenue}_t^{pool} = \frac{1}{\textsf{TVL}_t}\times\frac{f}{1+f}\ \sum_{\substack{\text{swaps $s$ on $pool$ in day $t$}}} \ p_{\tau_s}^{TI}(s)\cdot q^{TI}(s)
\end{align*}
where $\tau_s$ is the time of the swap and $p^{TI}_{\tau_s}(s)$ and $q^{TI}(s)$ are the dollar price and amount, respectively, of the token that swap $s$ puts \textit{into} the pool.\footnote{As fees on Uniswap are determined by the token-in amount, and our swap event tracker includes fees in $q^{TI}$, we multiply the sum across swaps by $f/(1+f)$.}

\paragraph{\textbf{Markout.}}
Markouts capture the informativeness of trades on an exchange by comparing the price of a trade to a benchmark price sometime after the trade, in our case the pool's mid-price, which indicates how favorable to the trader the swap was in hindsight.
Commonly used in traditional market microstructure, markouts have also been used as a proxy for LVR \cite{lehar2024fragmentation},\cite{milionis2022automated} in DEXs.

For each swap, we compare the swap price with the mid-price of the pool determined by the ``current tick'' at time $\tau_s+h_m$, where $\tau_s$ is the time of a swap $s$ and $h_m$ is the time horizon for computing markout. 
The resulting difference in price is then volume-weighted.
We aggregate markouts for all swaps in a given day and normalize by the end-of-day pool TVL:
\begin{align*}
    \textsf{Markout}_t^{pool} &= \frac{1}{\textsf{TVL}_t}\times  \sum_{\substack{\text{swaps $s$ on $pool$ in day $t$}}} D_s\cdot|q^{TI}(s)|\cdot\left(\left|\frac{q^{TO}(s)}{q^{TI}(s)}\right|-p^{pool}_{\tau_s+h_m}\right)
\end{align*}
where $D_s=1$ if the swapper is selling WETH and $D_s=-1$ otherwise (i.e.\ buying WETH), $q^{TO}(s)$ is the token-out amount for swap $s$, and $p^{pool}_{\tau_s+h_m}$ denotes the pool price, in units of the other token per WETH, at time $h_m$ after the swap occurred. 
Under this definition, more positive (negative) values indicate better (worse) LP profitability and thus less (more) adverse selection costs from swappers.

% \textcolor{blue}{should also consider to plot the distribution of the 5 min markout actual time, 1hr markout actual time, and half day markout actual time.}

% \subsection{Summary Statistics}

% \subsection{Data Visualizations}

\section{Methodology}

% Uniswap v3 introduced significant innovations to decentralized liquidity provision, most notably through concentrated liquidity. 
% Unlike traditional \amms or Uniswap v2, where liquidity is distributed evenly across all price ranges, Uniswap v3 allows liquidity providers to concentrate their liquidity within specific price bands. 
% This leads to more capital-efficient liquidity provision but also introduces complexities in understanding how liquidity dynamics influence key market outcomes, such as the bid-ask spread.

In AMMs with concentrated liquidity, market depth is not only influenced by the TVL in the pool, but is also by how concentrated that liquidity is across different price ranges. 
In this section, we introduce a novel method to measure concentration, which we use to distinguish between the effects of TVL and liquidity concentration on changes in effective spreads.
% By doing so, we can assess how much of the variation in spread is attributable to changes in token amounts versus changes in the distribution of liquidity across price ranges. 
This decomposition allows us to better understand the mechanics of liquidity provision in concentrated AMM pools and provide insights on LP behavior.

\subsection{Decomposing Spread in Uniswap v3}

The \textit{counterfactual v2 spread}, henceforth referred to as the \textsf{cfv2Spread}, is computed by looking at the TVL in a v3 pool at some given time, counterfactually considering a v2 pool with the same TVL and no trading fee such that the spot price on the v2 pool aligns with the CEX price at that time, and computing the effective spread as described in Section 3 on the counterfactual pool. 
Note that this v2 pool is counterfactual and does not correspond to the real v2 pool for the token pair. 
We align the counterfactual v2 pool's reserves with the CEX price to simulate the effect of arbitrageurs. % See Appendix \ref{appendix:summarystat} for time series plots of the counterfactual v2 spread.

As a v2 pool between tokens X and Y aligned to CEX prices has equal values of each token, the pool reserves $(X_t,Y_t)$ given TVL at time $t$ should satisfy
\begin{gather*}
    (X_t^{pool},Y_t^{pool}) = \left(\frac{\textsf{TVL}_t^{pool}}{2p_t^X},\frac{\textsf{TVL}_t^{pool}}{2p_t^Y}\right)
\end{gather*}
where $p_t^X$ and $p_t^Y$ are the CEX prices of tokens X and Y, respectively. A swap buying $\Delta_X$ tokens X for $\Delta_Y$ tokens Y must satisfy $(X-\Delta_X)(Y+\Delta_Y) = XY$.
Solving for $\Delta_Y$ and scaling by $\Delta_X$ yields the counterfactual ask price of
\begin{align*}
    A_t^{pool} = \frac{\Delta_Y}{\Delta_X} = \frac{Y_t^{pool}}{X_t^{pool}-\Delta_X}.
\end{align*}
A swap selling $\Delta_X$ tokens X for $\Delta_Y$ tokens Y must satisfy $(X+\Delta_X)(Y-\Delta_Y) = XY$.
Solving for $\Delta_Y$ and scaling by $\Delta_X$ yields the counterfactual bid price of
\begin{align*}
    B_t^{pool} = \frac{\Delta_Y}{\Delta_X} = \frac{Y_t^{pool}}{X_t^{pool}+\Delta_X}.
\end{align*}
The counterfactual v2 spread for a trade size of $\Delta$ WETH is then
\begin{align*}
    \textsf{cfv2Spread}_t^{pool} = 10^4\times \frac{A_t^{pool}-B_t^{pool}}{\frac{1}{2}(A_t^{pool}+B_t^{pool})} = 10^4\times\frac{4p_t^{ETH}}{\textsf{TVL}_t^{pool}}\Delta.
\end{align*}
The quotient between the actual v3 and counterfactual v2 spreads, given by
\begin{align*}
    \textsf{v3S/cfv2S} = \frac{\textsf{v3Spread}}{\textsf{cfv2Spread}},
\end{align*}
is a proxy for how well-concentrated the pool is around its mid-price: as \textsf{v3S/cfv2S} increases, the spread of the actual pool becomes higher relative to that of the counterfactual pool, meaning that liquidity is not well-concentrated in the actual pool; conversely, as \textsf{v3S/cfv2S} decreases, the spread of the actual pool becomes lower, suggesting a more efficient concentration of liquidity. 
Taking logarithms of the above equation reveals that
\begin{align*}
    \log\textsf{\,v3Spread} =  \log\textsf{\,cfv2Spread}+\log\textsf{\,v3S/cfv2S}.
\end{align*}

%%%%maybe like this?%%%
% This motivates the following three regression models:
% \begin{gather}
% y_{t+1}^{pool} = \beta_0
%                     + \beta_1\log\textsf{GasPrice}_t^{chain} 
%                     + \beta_2{\textsf{\,LogReturns}}_t^{pair}
%                     + \beta_3{\textsf{\,Volatility}}_t^{pair} \notag\\
%                     + \beta_4\log\textsf{FeeRevenue}_t^{pool} 
%                     + \beta_5{\textsf{\,Markout}}_t^{pool} + \gamma^\textit{pool} + \delta_t + \varepsilon^{pool}_{t+1}.
% \end{gather}
% where $y\in\{\log\textsf{v3Spread},\,\log\textsf{cfv2Spread},\log\textsf{v3S/cfv2S}\}$.

%%%%%%%%%

This motivates the following three regression models:
\begin{gather}
\log\textsf{v3Spread}_{t+1}^{pool} = \beta_0
                    + \beta_1\log\textsf{GasPrice}_t^{chain} 
                    + \beta_2{\textsf{\,LogReturns}}_t^{pair}
                    + \beta_3{\textsf{\,Volatility}}_t^{pair} \notag \\
                    + \beta_4\log\textsf{FeeRevenue}_t^{pool}
                    + \beta_5{\textsf{\,Markout}}_t^{pool} + \gamma^\textit{pool} + \delta_t + \varepsilon^{pool}_{t+1}
\label{equation1}
\end{gather}
\vspace{-2em}
\begin{gather}
\log\textsf{cfv2Spread}_{t+1}^{pool} = \beta_0
                    + \beta_1\log\textsf{GasPrice}_t^{chain} 
                    + \beta_2{\textsf{\,LogReturns}}_t^{pair}
                    + \beta_3{\textsf{\,Volatility}}_t^{pair} \notag \\
                    + \beta_4\log\textsf{FeeRevenue}_t^{pool}
                    + \beta_5{\textsf{\,Markout}}_t^{pool} + \gamma^\textit{pool} + \delta_t + \varepsilon^{pool}_{t+1}
\label{equation2}
\end{gather}
\vspace{-2em}
\begin{gather}
\log\textsf{v3S/cfv2S}_{t+1}^{pool} = \beta_0
                    + \beta_1\log\textsf{GasPrice}_t^{chain} 
                    + \beta_2{\textsf{\,LogReturns}}_t^{pair}
                    + \beta_3{\textsf{\,Volatility}}_t^{pair} \notag \\
                    + \beta_4\log\textsf{FeeRevenue}_t^{pool}
                    + \beta_5{\textsf{\,Markout}}_t^{pool} + \gamma^\textit{pool} + \delta_t + \varepsilon^{pool}_{t+1}
\label{equation3}
\end{gather}
Due to skewed data, we take logarithms of the spread, \textsf{GasPrice}, and fee revenue.
The terms $\gamma^\textit{pool}$ and $\delta_t$ represent pool-level and day-level fixed effects, respectively, while $\varepsilon^{pool}_{t+1}$ is the error term. 
The pool fixed effects capture time-invariant characteristics specific to each pool, such as whether the pool is included in the default Uniswap interface or other platform-specific settings that remain consistent.
Day fixed effects account for factors that affect all pools on a given date, such as regulatory news or shifts in overall market sentiment.
Standard errors are clustered at the pool level to account for heteroscedasticity and autocorrelation within pools over time. 

% where $\textsf{v3Spread}_{pool,t+1}$, $\textsf{cfv2Spread}_{pool,t+1}$, and $\textsf{v3S/cfv2S}_{pool,t+1}$ are liquidity metrics for a pool at time $t+1$, and $\textsf{GasPrice}_{chain,t}$, $\textsf{LogReturns}_t^{pair}$, $\textsf{Volatility}_t^{pair}$, $\textsf{FeeRevenue}_t^{pool}$, and $\textsf{Markout}_t^{pool}$ for the pool at time $t$.

\subsection{Results and Discussion}

We estimate the regression models with a trade size of $\Delta=1$ WETH to compute spreads, a return horizon $h_r$ of 1 day, and a markout horizon $h_m$ of 5 minutes.\footnote{The \textsf{GasPrice} and markout variable exhibits extreme values that could disproportionately influence the regression results. 
To address this, we exclude pool-days where the \textsf{GasPrice} markout exceeds 5 standard deviations from their respective means, removing 56 pool-days from the sample.} 
% Regression results using the full sample are provided in the Appendix.}
Prior to estimation, we normalize each independent variable in the data matrix to have mean zero and standard deviation one, preserving the significance of the coefficients while allowing for interpretable effect sizes.\footnote{Our main results are generally robust to modifications in $\Delta$, $h_r$ and $h_m$. Specifically, we also have considered $\Delta\in\{0.1,10\}$, $h_r=7$ days, and $h_m=1$ hour.}

Note that our decomposition implies that for each control, the estimated coefficients from model (\ref{equation2}) and (\ref{equation3}) sum to the coefficient from estimating model (\ref{equation1}), though their significance levels may vary:
\begin{align*}
    \beta_i^{\log\textsf{v3Spread}} = \beta_i^{\log\textsf{cfv2Spread}} +\beta_i^{\log\textsf{v3S/cfv2S}}\,\,\,\forall i.
\end{align*}
We also regress the log-TVL at time $t+1$ on the sets of independent variables at time $t$ for completeness and interpretability. 
Since the counterfactual v2 spread is a function of price and TVL, and fixed effects by pool and day are included, the coefficients from this regression will equal to those from regression (\ref{equation2}) times minus one:
\begin{align*}
    \beta_i^{\log\textsf{cfv2Spread}} = -\beta_i^{\log\textsf{TVL}}\, \forall i.
\end{align*}

\setlength{\tabcolsep}{0em} % for the horizontal padding
\renewcommand{\arraystretch}{1}% for the vertical padding

\begin{table}[H] \centering
\caption{
Baseline Regression Model (\ref{equation1}) with Decomposition (\ref{equation2}) $+$ (\ref{equation3})
}
\begin{tabular}{l 
>{\centering\arraybackslash}p{0.2\linewidth} 
>{\centering\arraybackslash}p{0.2\linewidth} 
>{\centering\arraybackslash}p{0.2\linewidth} 
>{\centering\arraybackslash}p{0.2\linewidth}}
\toprule
& (1) & (2) & (3) & (4) \\
& $\log\textsf{v3Spread}$ & $\log\textsf{cfv2Spread}$ & $\log\textsf{v3S/cfv2S}$ & $\log\textsf{TVL}$ \\
\midrule
% const & 1.383$^{***}$ & 3.016$^{***}$ & -1.633$^{***}$ & 15.257$^{***}$ \\
% & (0.000) & (0.000) & (0.000) & (0.000) \\
 $\log\textsf{GasPrice}$ & 0.213$^{}$ & 0.085$^{}$ & 0.128$^{***}$ & -0.085$^{}$ \\
& (0.132) & (0.126) & (0.048) & (0.126) \\
 \textsf{LogReturns} & -0.033$^{***}$ & -0.009$^{}$ & -0.024$^{***}$ & 0.009$^{}$ \\
& (0.008) & (0.006) & (0.005) & (0.006) \\
 \textsf{Volatility} & 0.401$^{***}$ & 0.101$^{**}$ & 0.300$^{***}$ & -0.101$^{**}$ \\
& (0.053) & (0.044) & (0.027) & (0.044) \\
 $\log\textsf{FeeRevenue}$ & -0.928$^{***}$ & -0.237$^{***}$ & -0.690$^{***}$ & 0.237$^{***}$ \\
& (0.117) & (0.074) & (0.086) & (0.074) \\
 \textsf{Markout} & -0.086$^{***}$ & -0.169$^{***}$ & 0.083$^{***}$ & 0.169$^{***}$ \\
& (0.028) & (0.019) & (0.021) & (0.019) \\
\hline \\[-1.8ex]
 Observations & 38440 & 38440 & 38440 & 38440 \\
 N. of groups & 40 & 40 & 40 & 40 \\
 $R^2$ & 0.313 & 0.078 & 0.435 & 0.078 \\
\bottomrule \\[-2.25ex]
\multicolumn{5}{l}{Pool and day fixed effects are included; standard errors are clustered at the pool level.} \\
\textit{Note:} & \multicolumn{4}{r}{$^{*}$p$<$0.1; $^{**}$p$<$0.05; $^{***}$p$<$0.01}
\end{tabular}
\label{table:baseline-usethis}
\end{table}

\vspace{-1.5em}

\autoref{table:baseline-usethis} presents the estimated effects of factors on each dependent variable specification. 
We omit $\beta_0$ estimates for brevity.
All factors except for \textsf{GasPrice} significantly impact the overall effective spread. 
Specifically, v3 spreads are increasing in volatility and is decreasing in returns, fee revenue, and markout.

% We now turn to the two channels that determine spreads.

The estimated coefficients on volatility, fee revenue, and markout for regression (2) are significant, suggesting that these factors have predictive power on how TVL affects overall spreads. Higher fee revenue and better markouts against swappers indicate more profitability for LPs, incentivizing them to provide liquidity, thus reducing spreads. Since volatility is associated with LP losses (both impermanent loss and LVR), higher volatility lowers liquidity provision.

All factors are significant in regression (3), implying that they play an important role in liquidity concentration. Higher \textsf{GasPrice}s increase rebalancing costs, leading to ``stale'' positions that are not concentrated around current pool prices. 
More volatility and negative markout (indicating informed trading) lead to LPs widening their price ranges in order to, as explained in \cite{capponi2024liquidity}, create a more convex pricing function that reduces losses to informed traders and volatile prices. Conversely, more fee revenue means that LPs can increase their profits by targeting narrower price ranges with a larger concentration liquidity, according to \cite{cartea2024v3}. Note that putting a fixed amount of assets in a wider range lowers the amount of ``virtual liquidity'' (see \cite{adams2021uniswap}) in the AMM, resulting in lower spreads.

As for returns, we note that in traditional markets, liquidity tends to dry up during market declines and periods of increased volatility \cite{chordia2001market,pastor2003liquidity}.
This suggests that higher returns should positively predict market depth, while higher volatility has a negative effect.
These are consistent with our results, suggesting that this stylized fact also carries over to decentralized markets when WETH is used as a numeraire.

\section{Extension: External Liquidity}

% \subsection{Impact of External Liquidity on Spread}

The success of AMMs has lead to the proliferation of DEXs, with there being over one hundred DEXs at the time of writing. 
The increase in competition between DEXs introduces challenges such as the fragmentation of liquidity across multiple liquidity pools and DEXs. 
In addition, the introduction of DEX aggregators has led to new methods for order routing, leveraging liquidity from various on-chain sources and off-chain private market makers (PMMs). While these services improve outcomes for swappers \cite{bachu2024quantifying}, PMMs earn fees that would have otherwise gone to on-chain LPs, affecting their overall profitability.

\subsection{Measuring External Liquidity}

To better understand how these external liquidity sources might affect on-chain liquidity provision, we introduce variables that (i) capture the volume of swaps taking place on other DEXs and (ii) filled by private liquidity due to aggregator routing.
Using data from Dune Analytics, we track swap volume on other DEXs and routed through aggregators, isolating swaps filled completely by private liquidity by comparing transaction hashes with on-chain events. 
A simple heuristic to identify these types of swaps is to take all swaps emitting events to aggregator trackers that did not emit a swap event to on-chain data trackers. 
% From this, we compute the \textit{DEX competitor trading volume share} and \textit{internalization ratio}, defined as follows.

\paragraph{\textbf{Competitor Market Share}.} 
For a set $\mathcal{D}$ of DEXs (including the given DEX) and a chain-pair, we compute the fraction of trading volume for that chain-pair occurring outside of the given DEX on day $t$:\footnote{Typically, market share for DEXs is evaluated in terms of trading volume. As a single DEX can have several pools trading a pair on a chain, we require a platform-level metric to assess competition and trader's sentiments towards a given DEX.}
\begin{align*}
    \textsf{CompetitorShare}_t^{chain,pair} = 1-\frac{v_t}{\sum_{D\in\mathcal{D}}v_t^D}
\end{align*}
where $v_t$ and $v_t^D$ are the swap volumes on the given DEX and DEX $D$ in USD, respectively, for a given chain-pair on day $t$. 

\paragraph{\textbf{Internalization Ratio.}}
Given a set $\mathcal{D}$ of DEXs, a set $\mathcal{A}$ of aggregators, and a chain-pair, we compute the proportion of swap volume routed or internalized by private market makers active on $\mathcal{A}$, henceforth referred to as ``private volume,'' to the total on-chain plus private volume for that chain-pair on day $t$:\footnote{Since an aggregator could route trades to a variety of DEXs, we need $\mathcal{D}$ to include all DEX that the aggregators in $\mathcal{A}$ may route to.}
\begin{align*}
    \textsf{Internalization}_t^{chain,pair} = \frac{\sum_{A\in\mathcal{A}}v_t^A}{\sum_{D\in\mathcal{D}}v_t^D+\sum_{A\in\mathcal{A}}v_t^A}
\end{align*}
where $v_t^A$ is the private volume on aggregator $A$ and $v_t^D$ is the swap volume on DEX $D$, both in USD, for a given chain-pair on day $t$. 

\paragraph{\textbf{Extended Regression Model.}}
We take $\mathcal{D}$ as the 130 DEXs whose swap events are tracked on Dune Analytics and $\mathcal{A}$ as the 13 aggregators whose swap events are tracked on Dune Analytics.
We add the competitor market share and internalization ratio variables to the baseline model and estimate the model with the dependent variable specifications in models (\ref{equation1})--(\ref{equation3}):
% We take $\mathcal{D}$ to be the set of all 130 DEXs whose swap events are tracked on Dune Analytics. 
% Since we only include Uniswap liquidity pools in our sample, the competitor market share is simply the ratio of Uniswap volume to total DEX volume for a given chain-pair in our case. 
% We consider two different aggregator sets $\mathcal{A}$, one that includes all 13 aggregators whose swaps are tracked on Dune Analytics, and another that only contains UniswapX. \footnote{Swaps on UniswapX filled by private liquidity are not tracked on Dune.}
% We add these variables to the baseline model and estimate the model using the dependent variable specifications in models (\ref{equation1})--(\ref{equation3}):
%We then posit the following regression framework adapted from \autoref{equation1}. Further more, similar to the previous section, we also analysis the decomposition like in \autoref{equation2} and \autoref{equation3} with different dependent variable specification.
\begin{align}
y_{t+1}^{pool} &= \beta_0 
                    + \beta_1 \log\textsf{GasPrice}_t^{chain} 
                    + \beta_2\textsf{\,LogReturns}_t^{pair} 
                    + \beta_3{\textsf{\,Volatility}}_t^{pair} \notag \\
                    & + \beta_4 \log\textsf{FeeRevenue}_t^{pool}
                    + \beta_5{\textsf{\,Markout}}_t^{pool} 
                    + \beta_6{\textsf{\,CompetitorShare}}_t^{chain,pair} \notag \\       
                    & + \beta_7{\textsf{\,Internalization}}_t^{chain,pair} 
                    + \gamma^\textit{pool} + \delta_t + \varepsilon_{t+1}^{pool}
\label{equation4}
\end{align}
where $y\in\{\log\textsf{v3Spread},\,\log\textsf{cfv2Spread},\log\textsf{v3S/cfv2S}\}$.\footnote{We performed robustness checks similar to those in footnote 4.}

\subsection{Results and Discussion}

\autoref{table:private-all} displays the estimation results of each dependent variable using the extended model that includes the competitor trading volume share and internalization ratios.
We find that a higher competitor share of the token pair predicts higher effective spreads, while there is no significant explanatory power from internalization.
Interestingly, the channels in which competitor share and internalization affect market depth differ: the former affects liquidity via concentration while the latter affects liquidity through value locked.

One possible explanation for this difference is that while providing liquidity privately and to competing DEXs are both alternative opportunities for LPs to earn fee revenue, providing liquidity privately is more discretionary, as PMMs can choose which orders to fill, while providing liquidity to a competing DEX requires the LP to take the opposite position of all trades routed to the DEX.
The greater risk involved in this option incentivizes LPs to widen price ranges, following the same intuition as the discussion on volatility and markout. 
Conversely, the lesser risks involved in being a PMM mean that LPs choosing this option may not need to widen ranges on existing position, instead directly withdrawing liquidity from pools to serve as private liquidity.

\vspace{-1.5em}

\begin{table}[H] \centering
\caption{
Model (\ref{equation4}) with External Liquidity Variables}
\begin{tabular}{l 
>{\centering\arraybackslash}p{0.2\linewidth} 
>{\centering\arraybackslash}p{0.2\linewidth} 
>{\centering\arraybackslash}p{0.2\linewidth} 
>{\centering\arraybackslash}p{0.2\linewidth}}
\toprule
& (1) & (2) & (3) & (4) \\
& $\log\textsf{v3Spread}$ & $\log\textsf{cfv2Spread}$ & $\log\textsf{v3S/cfv2S}$ & $\log\textsf{TVL}$ \\
\midrule
% const & 1.383$^{***}$ & 3.016$^{***}$ & -1.633$^{***}$ & 15.257$^{***}$ \\
% & (0.000) & (0.000) & (0.000) & (0.000) \\
 $\log\textsf{GasPrice}$ & 0.178$^{}$ & 0.083$^{}$ & 0.095$^{**}$ & -0.083$^{}$ \\
& (0.126) & (0.120) & (0.047) & (0.120) \\
 \textsf{LogReturns} & -0.033$^{***}$ & -0.009$^{}$ & -0.024$^{***}$ & 0.009$^{}$ \\
& (0.008) & (0.006) & (0.006) & (0.006) \\
 \textsf{Volatility} & 0.379$^{***}$ & 0.089$^{**}$ & 0.290$^{***}$ & -0.089$^{**}$ \\
& (0.052) & (0.045) & (0.026) & (0.045) \\
 $\log\textsf{FeeRevenue}$ & -0.869$^{***}$ & -0.201$^{**}$ & -0.668$^{***}$ & 0.201$^{**}$ \\
& (0.119) & (0.079) & (0.079) & (0.079) \\
 \textsf{Markout} & -0.088$^{***}$ & -0.169$^{***}$ & 0.081$^{***}$ & 0.169$^{***}$ \\
& (0.026) & (0.019) & (0.020) & (0.019) \\
 \textsf{CompetitorShare} & 0.222$^{***}$ & 0.088$^{}$ & 0.134$^{***}$ & -0.088$^{}$ \\
& (0.065) & (0.056) & (0.031) & (0.056) \\
 \textsf{Internalization} & 0.062$^{}$ & 0.113$^{***}$ & -0.051$^{}$ & -0.113$^{***}$ \\
& (0.082) & (0.027) & (0.070) & (0.027) \\
\hline \\[-1.8ex]
 Observations & 38440 & 38440 & 38440 & 38440 \\
 N. of groups & 40 & 40 & 40 & 40 \\
 $R^2$ & 0.327 & 0.093 & 0.447 & 0.093 \\
\bottomrule \\[-2.25ex]
\multicolumn{5}{l}{Pool and day fixed effects are included; standard errors are clustered at the pool level.} \\
\textit{Note:} & \multicolumn{4}{r}{$^{*}$p$<$0.1; $^{**}$p$<$0.05; $^{***}$p$<$0.01}
\end{tabular}
\label{table:private-all}
\end{table}

\section{Conclusion}

This study provides a valuable understanding of what drives liquidity on \dexs, specifically within the Uniswap v3 protocol, having analyzed various factors and their explanatory power in predicting future market depth.
We introduced the v2 counterfactual spread metric to decompose the drivers of overall effective spread, distinguishing between impacts through TVL and liquidity concentration. 
Our findings suggest that increased competition between DEXs and the presence of private liquidity sources are significant contributors to liquidity fragmentation on Uniswap v3, though they influence market depth via differing channels. % , as external factors primarily influence effective spreads through TVL reduction.

Our findings have significant implications for both LPs and \dex designers.
Understanding these dynamics is essential for LPs looking to optimize liquidity provision strategies, and for \dex designers, these insights can guide the development of features that address the adverse effects of liquidity fragmentation.
Our results on private liquidity are optimistic for the coexistence of DEX aggregators and on-chain liquidity, as internalization shows no significant effect on \textit{overall} market depth.

Future research could explore more elements of competition and internalization in multi-DEX ecosystems, especially as aggregator services evolve. 
Further studies on how alternative blockchain environments and emerging Layer 2 solutions support affect liquidity provision can provide a broader perspective on the scalability and sustainability of DEXs in a rapidly growing DeFi landscape.

\begin{credits}
\subsubsection{Acknowledgements and Disclosures.}
The fifth author was supported by the Briger Family Digital Finance Lab at Columbia Business School and is
an advisor to fintech companies. 
\end{credits}

%\clearpage

\bibliographystyle{splncs04}
\bibliography{reference}

\clearpage

\appendix
\section{Data Visualizations}
\label{sec:datavisualization}

\vspace{-1em}

\begin{figure}[H]
    \centering
    \hspace{2em}
    \includegraphics[scale=0.38]{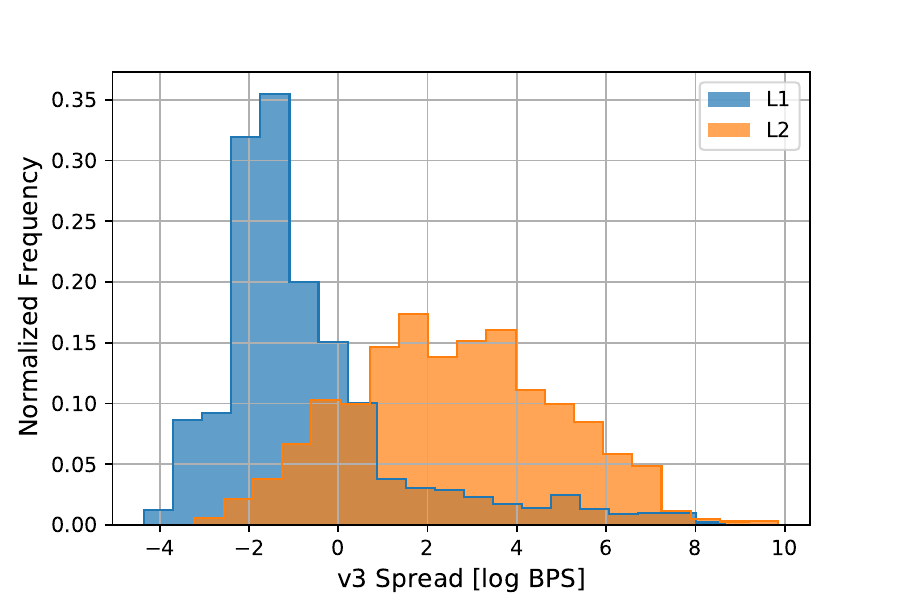}
    \hspace{-2em}
    \includegraphics[scale=0.38]{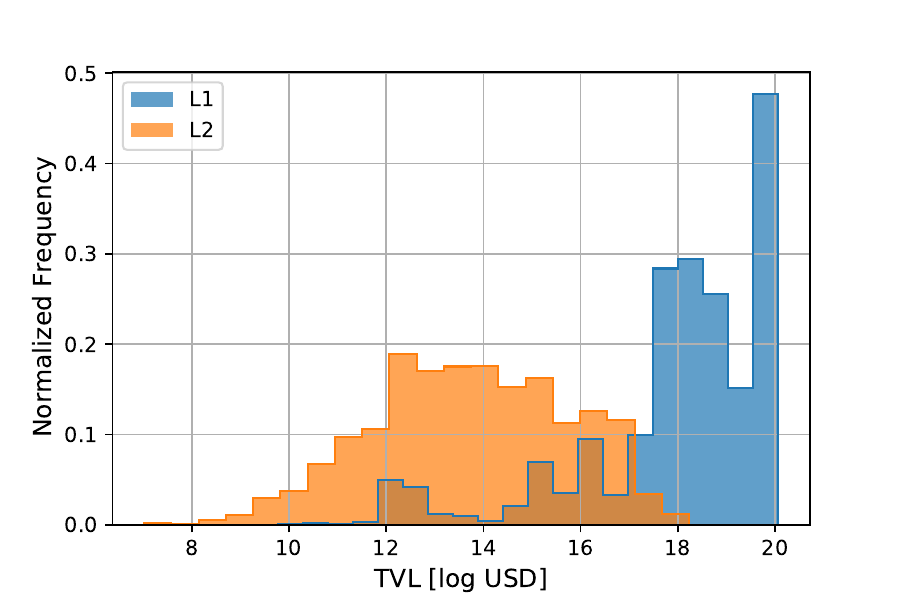}
    \caption{\centering Normalized histograms of effective spreads for $\Delta=1$ WETH (left) and TVL (right), separated by L1 and L2 networks.}
    \label{fig:liquidity1}
\end{figure}

\vspace{-2em}

\begin{figure}[H]
    \centering
    \includegraphics[scale=0.5]{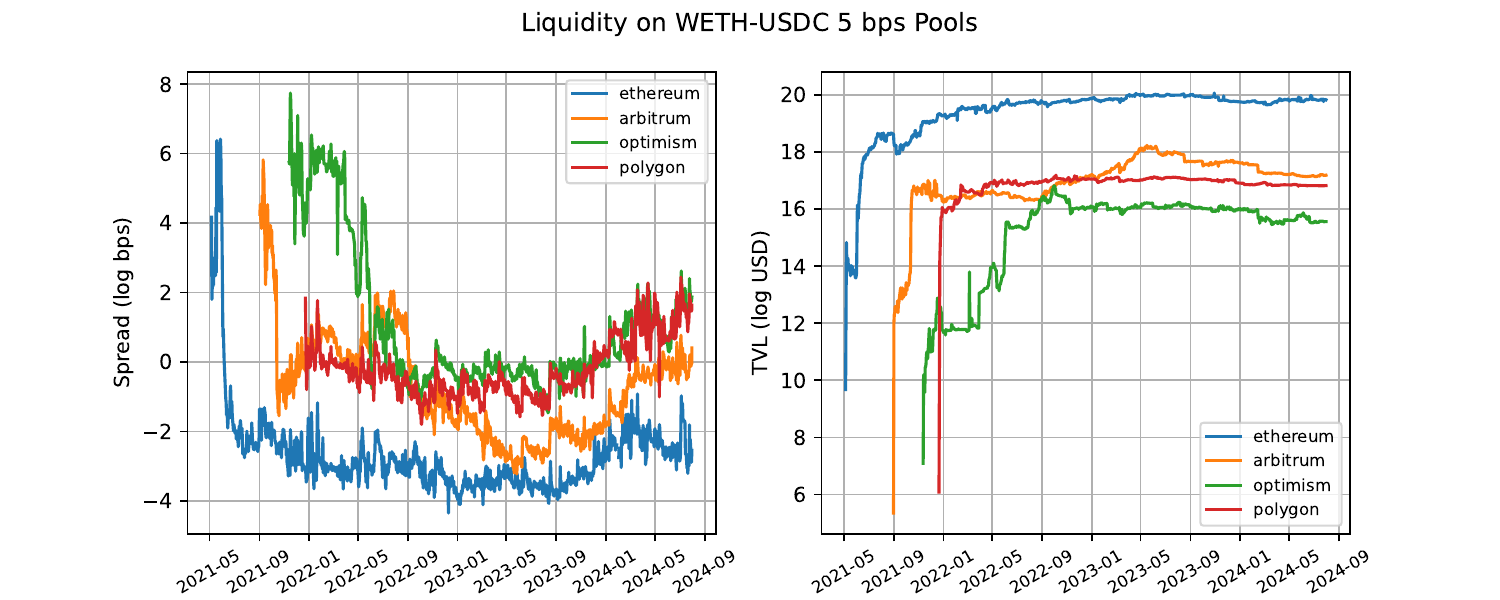} \\[0.5\baselineskip]
    \includegraphics[scale=0.5]{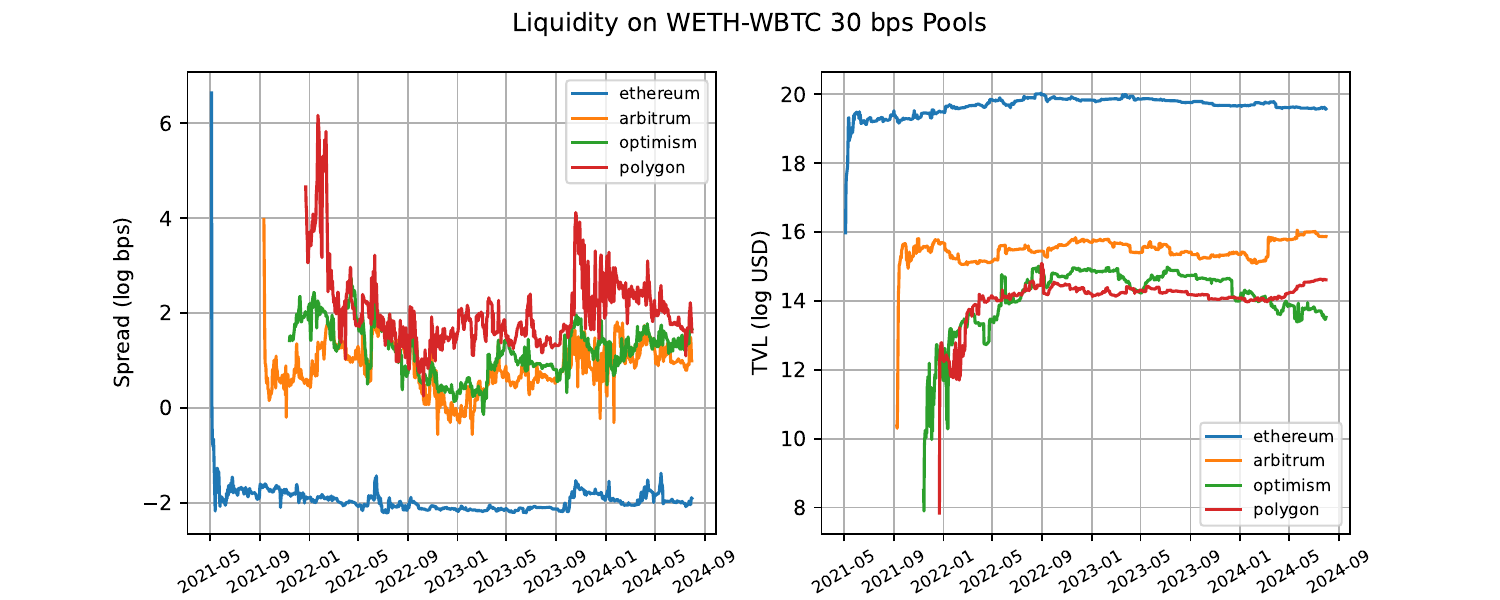}
    \caption{\centering Time series of effective spreads for $\Delta=1$ WETH (left) and TVL (right) for selected liquidity pools.}
    \label{fig:liquidity1}
\end{figure}

\clearpage
\section{Robustness Checks}

\setlength{\tabcolsep}{0em} % for the horizontal padding
\renewcommand{\arraystretch}{1}% for the vertical padding

\subsection{Weekly Returns}

\vspace{-2em}

\begin{table}[H] \centering
\caption{
Baseline Regression Model ($\Delta=1$, $h_r=7$ days, $h_m=5$ mins)
}
\begin{tabular}{l 
>{\centering\arraybackslash}p{0.2\linewidth} 
>{\centering\arraybackslash}p{0.2\linewidth} 
>{\centering\arraybackslash}p{0.2\linewidth} 
>{\centering\arraybackslash}p{0.2\linewidth}}
\toprule
& (1) & (2) & (3) & (4) \\
& $\log\textsf{v3Spread}$ & $\log\textsf{cfv2Spread}$ & $\log\textsf{v3S/cfv2S}$ & $\log\textsf{TVL}$ \\
\midrule
% const & 1.383$^{***}$ & 3.016$^{***}$ & -1.633$^{***}$ & 15.257$^{***}$ \\
% & (0.000) & (0.000) & (0.000) & (0.000) \\
 $\log\textsf{GasPrice}$ & 0.213$^{}$ & 0.085$^{}$ & 0.128$^{***}$ & -0.085$^{}$ \\
& (0.132) & (0.125) & (0.048) & (0.125) \\
 \textsf{LogReturns} & -0.051$^{***}$ & -0.024$^{}$ & -0.027$^{***}$ & 0.024$^{}$ \\
& (0.016) & (0.015) & (0.009) & (0.015) \\
 \textsf{Volatility} & 0.386$^{***}$ & 0.092$^{**}$ & 0.294$^{***}$ & -0.092$^{**}$ \\
& (0.053) & (0.044) & (0.026) & (0.044) \\
 $\log\textsf{FeeRevenue}$ & -0.930$^{***}$ & -0.239$^{***}$ & -0.691$^{***}$ & 0.239$^{***}$ \\
& (0.117) & (0.074) & (0.086) & (0.074) \\
 \textsf{Markout} & -0.086$^{***}$ & -0.169$^{***}$ & 0.083$^{***}$ & 0.169$^{***}$ \\
& (0.028) & (0.019) & (0.021) & (0.019) \\
\hline \\[-1.8ex]
 Observations & 38440 & 38440 & 38440 & 38440 \\
 N. of groups & 40 & 40 & 40 & 40 \\
 $R^2$ & 0.314 & 0.079 & 0.435 & 0.079 \\
\bottomrule \\[-2.25ex]
\multicolumn{5}{l}{Pool and day fixed effects are included; standard errors are clustered at the pool level.} \\
\textit{Note:} & \multicolumn{4}{r}{$^{*}$p$<$0.1; $^{**}$p$<$0.05; $^{***}$p$<$0.01}
\end{tabular}
\label{table:baseline-usethis}
\end{table}

\vspace{-4em}

\begin{table}[H] \centering
\caption{
Extended Regression Model ($\Delta=1$, $h_r=7$ days, $h_m=5$ mins)
}
\begin{tabular}{l 
>{\centering\arraybackslash}p{0.2\linewidth} 
>{\centering\arraybackslash}p{0.2\linewidth} 
>{\centering\arraybackslash}p{0.2\linewidth} 
>{\centering\arraybackslash}p{0.2\linewidth}}
\toprule
& (1) & (2) & (3) & (4) \\
& $\log\textsf{v3Spread}$ & $\log\textsf{cfv2Spread}$ & $\log\textsf{v3S/cfv2S}$ & $\log\textsf{TVL}$ \\
\midrule
% const & 1.383$^{***}$ & 3.016$^{***}$ & -1.633$^{***}$ & 15.257$^{***}$ \\
% & (0.000) & (0.000) & (0.000) & (0.000) \\
 $\log\textsf{GasPrice}$ & 0.177$^{}$ & 0.083$^{}$ & 0.095$^{**}$ & -0.083$^{}$ \\
& (0.126) & (0.120) & (0.047) & (0.120) \\
 \textsf{LogReturns} & -0.050$^{***}$ & -0.024$^{}$ & -0.027$^{***}$ & 0.024$^{}$ \\
& (0.016) & (0.015) & (0.009) & (0.015) \\
 \textsf{Volatility} & 0.364$^{***}$ & 0.080$^{*}$ & 0.284$^{***}$ & -0.080$^{*}$ \\
& (0.052) & (0.045) & (0.025) & (0.045) \\
 $\log\textsf{FeeRevenue}$ & -0.872$^{***}$ & -0.202$^{**}$ & -0.669$^{***}$ & 0.202$^{**}$ \\
& (0.119) & (0.079) & (0.079) & (0.079) \\
 \textsf{Markout} & -0.088$^{***}$ & -0.169$^{***}$ & 0.081$^{***}$ & 0.169$^{***}$ \\
& (0.026) & (0.019) & (0.020) & (0.019) \\
 \textsf{CompetitorShare} & 0.222$^{***}$ & 0.088$^{}$ & 0.134$^{***}$ & -0.088$^{}$ \\
& (0.065) & (0.056) & (0.031) & (0.056) \\
 \textsf{Internalization} & 0.061$^{}$ & 0.113$^{***}$ & -0.051$^{}$ & -0.113$^{***}$ \\
& (0.082) & (0.027) & (0.070) & (0.027) \\
\hline \\[-1.8ex]
 Observations & 38440 & 38440 & 38440 & 38440 \\
 N. of groups & 40 & 40 & 40 & 40 \\
 $R^2$ & 0.327 & 0.094 & 0.447 & 0.094 \\
\bottomrule \\[-2.25ex]
\multicolumn{5}{l}{Pool and day fixed effects are included; standard errors are clustered at the pool level.} \\
\textit{Note:} & \multicolumn{4}{r}{$^{*}$p$<$0.1; $^{**}$p$<$0.05; $^{***}$p$<$0.01}
\end{tabular}
\label{table:private-all}
\end{table}

\subsection{Hourly \textsf{Markout}}

\vspace{-2em}

\begin{table}[H] \centering
\caption{
Baseline Regression Model ($\Delta=1$, $h_r=1$ day, $h_m=1$ hour)
}
\begin{tabular}{l 
>{\centering\arraybackslash}p{0.2\linewidth} 
>{\centering\arraybackslash}p{0.2\linewidth} 
>{\centering\arraybackslash}p{0.2\linewidth} 
>{\centering\arraybackslash}p{0.2\linewidth}}
\toprule
& (1) & (2) & (3) & (4) \\
& $\log\textsf{v3Spread}$ & $\log\textsf{cfv2Spread}$ & $\log\textsf{v3S/cfv2S}$ & $\log\textsf{TVL}$ \\
\midrule
% const & 1.383$^{***}$ & 3.016$^{***}$ & -1.633$^{***}$ & 15.257$^{***}$ \\
% & (0.000) & (0.000) & (0.000) & (0.000) \\
 $\log\textsf{GasPrice}$ & 0.206$^{}$ & 0.076$^{}$ & 0.129$^{***}$ & -0.076$^{}$ \\
& (0.133) & (0.127) & (0.049) & (0.127) \\
 \textsf{LogReturns} & -0.033$^{***}$ & -0.009$^{}$ & -0.024$^{***}$ & 0.009$^{}$ \\
& (0.008) & (0.006) & (0.006) & (0.006) \\
 \textsf{Volatility} & 0.402$^{***}$ & 0.103$^{**}$ & 0.298$^{***}$ & -0.103$^{**}$ \\
& (0.054) & (0.045) & (0.027) & (0.045) \\
 $\log\textsf{FeeRevenue}$ & -0.918$^{***}$ & -0.223$^{***}$ & -0.695$^{***}$ & 0.223$^{***}$ \\
& (0.117) & (0.075) & (0.086) & (0.075) \\
 \textsf{Markout} & -0.065$^{***}$ & -0.144$^{***}$ & 0.079$^{***}$ & 0.144$^{***}$ \\
& (0.017) & (0.017) & (0.020) & (0.017) \\
\hline \\[-1.8ex]
 Observations & 38447 & 38447 & 38447 & 38447 \\
 N. of groups & 40 & 40 & 40 & 40 \\
 $R^2$ & 0.309 & 0.065 & 0.435 & 0.065 \\
\bottomrule \\[-2.25ex]
\multicolumn{5}{l}{Pool and day fixed effects are included; standard errors are clustered at the pool level.} \\
\textit{Note:} & \multicolumn{4}{r}{$^{*}$p$<$0.1; $^{**}$p$<$0.05; $^{***}$p$<$0.01}
\end{tabular}
\label{table:baseline-usethis}
\end{table}

\vspace{-4em}

\begin{table}[H] \centering
\caption{
Extended Regression Model ($\Delta=1$, $h_r=1$ day, $h_m=1$ hour)
}
\begin{tabular}{l 
>{\centering\arraybackslash}p{0.2\linewidth} 
>{\centering\arraybackslash}p{0.2\linewidth} 
>{\centering\arraybackslash}p{0.2\linewidth} 
>{\centering\arraybackslash}p{0.2\linewidth}}
\toprule
& (1) & (2) & (3) & (4) \\
& $\log\textsf{v3Spread}$ & $\log\textsf{cfv2Spread}$ & $\log\textsf{v3S/cfv2S}$ & $\log\textsf{TVL}$ \\
\midrule
% const & 1.383$^{***}$ & 3.016$^{***}$ & -1.633$^{***}$ & 15.257$^{***}$ \\
% & (0.000) & (0.000) & (0.000) & (0.000) \\
 $\log\textsf{GasPrice}$ & 0.170$^{}$ & 0.074$^{}$ & 0.096$^{**}$ & -0.074$^{}$ \\
& (0.127) & (0.121) & (0.047) & (0.121) \\
 \textsf{LogReturns} & -0.033$^{***}$ & -0.009$^{}$ & -0.024$^{***}$ & 0.009$^{}$ \\
& (0.008) & (0.006) & (0.006) & (0.006) \\
 \textsf{Volatility} & 0.380$^{***}$ & 0.091$^{**}$ & 0.288$^{***}$ & -0.092$^{**}$ \\
& (0.052) & (0.046) & (0.026) & (0.046) \\
 $\log\textsf{FeeRevenue}$ & -0.859$^{***}$ & -0.186$^{**}$ & -0.673$^{***}$ & 0.186$^{**}$ \\
& (0.119) & (0.080) & (0.078) & (0.080) \\
 \textsf{Markout} & -0.067$^{***}$ & -0.144$^{***}$ & 0.078$^{***}$ & 0.144$^{***}$ \\
& (0.016) & (0.018) & (0.019) & (0.018) \\
 \textsf{CompetitorShare} & 0.223$^{***}$ & 0.089$^{}$ & 0.134$^{***}$ & -0.089$^{}$ \\
& (0.065) & (0.057) & (0.032) & (0.057) \\
 \textsf{Internalization} & 0.063$^{}$ & 0.114$^{***}$ & -0.051$^{}$ & -0.114$^{***}$ \\
& (0.082) & (0.027) & (0.070) & (0.027) \\
\hline \\[-1.8ex]
 Observations & 38447 & 38447 & 38447 & 38447 \\
 N. of groups & 40 & 40 & 40 & 40 \\
 $R^2$ & 0.323 & 0.080 & 0.447 & 0.080 \\
\bottomrule \\[-2.25ex]
\multicolumn{5}{l}{Pool and day fixed effects are included; standard errors are clustered at the pool level.} \\
\textit{Note:} & \multicolumn{4}{r}{$^{*}$p$<$0.1; $^{**}$p$<$0.05; $^{***}$p$<$0.01}
\end{tabular}
\label{table:private-all}
\end{table}

\subsection{Small Trade Size (0.1 WETH)}

\vspace{-2em}

\begin{table}[H] \centering
\caption{Baseline Regression Model ($\Delta=0.1$, $h_r=1$ day, $h_m=5$ mins)}
\begin{tabular}{l 
>{\centering\arraybackslash}p{0.2\linewidth} 
>{\centering\arraybackslash}p{0.2\linewidth} 
>{\centering\arraybackslash}p{0.2\linewidth} 
>{\centering\arraybackslash}p{0.2\linewidth}}
\toprule
& (1) & (2) & (3) & (4) \\
& $\log\textsf{v3Spread}$ & $\log\textsf{cfv2Spread}$ & $\log\textsf{v3S/cfv2S}$ & $\log\textsf{TVL}$ \\
\midrule
% const & 1.383$^{***}$ & 3.016$^{***}$ & -1.633$^{***}$ & 15.257$^{***}$ \\
% & (0.000) & (0.000) & (0.000) & (0.000) \\
 $\log\textsf{GasPrice}$ & 0.213$^{}$ & 0.085$^{}$ & 0.128$^{***}$ & -0.085$^{}$ \\
& (0.132) & (0.126) & (0.048) & (0.126) \\
 \textsf{LogReturns} & -0.033$^{***}$ & -0.009$^{}$ & -0.024$^{***}$ & 0.009$^{}$ \\
& (0.008) & (0.006) & (0.005) & (0.006) \\
 \textsf{Volatility} & 0.401$^{***}$ & 0.101$^{**}$ & 0.300$^{***}$ & -0.101$^{**}$ \\
& (0.053) & (0.044) & (0.027) & (0.044) \\
 $\log\textsf{FeeRevenue}$ & -0.928$^{***}$ & -0.237$^{***}$ & -0.690$^{***}$ & 0.237$^{***}$ \\
& (0.117) & (0.074) & (0.086) & (0.074) \\
 \textsf{Markout} & -0.086$^{***}$ & -0.169$^{***}$ & 0.083$^{***}$ & 0.169$^{***}$ \\
& (0.028) & (0.019) & (0.021) & (0.019) \\
\hline \\[-1.8ex]
 Observations & 38440 & 38440 & 38440 & 38440 \\
 N. of groups & 40 & 40 & 40 & 40 \\
 $R^2$ & 0.313 & 0.078 & 0.435 & 0.078 \\
\bottomrule \\[-2.25ex]
\multicolumn{5}{l}{Pool and day fixed effects are included; standard errors are clustered at the pool level.} \\
\textit{Note:} & \multicolumn{4}{r}{$^{*}$p$<$0.1; $^{**}$p$<$0.05; $^{***}$p$<$0.01}
\end{tabular}
\label{table:baseline-usethis}
\end{table}

\vspace{-4em}

\begin{table}[H] \centering
\caption{
Extended Regression Model ($\Delta=0.1$, $h_r=1$ day, $h_m=5$ mins)
}
\begin{tabular}{l 
>{\centering\arraybackslash}p{0.2\linewidth} 
>{\centering\arraybackslash}p{0.2\linewidth} 
>{\centering\arraybackslash}p{0.2\linewidth} 
>{\centering\arraybackslash}p{0.2\linewidth}}
\toprule
& (1) & (2) & (3) & (4) \\
& $\log\textsf{v3Spread}$ & $\log\textsf{cfv2Spread}$ & $\log\textsf{v3S/cfv2S}$ & $\log\textsf{TVL}$ \\
\midrule
% const & 1.383$^{***}$ & 3.016$^{***}$ & -1.633$^{***}$ & 15.257$^{***}$ \\
% & (0.000) & (0.000) & (0.000) & (0.000) \\
 $\log\textsf{GasPrice}$ & 0.093$^{}$ & 0.035$^{}$ & 0.058$^{}$ & -0.035$^{}$ \\
& (0.144) & (0.130) & (0.059) & (0.130) \\
 \textsf{LogReturns} & -0.041$^{***}$ & -0.017$^{***}$ & -0.024$^{***}$ & 0.017$^{***}$ \\
& (0.008) & (0.006) & (0.006) & (0.006) \\
 \textsf{Volatility} & 0.407$^{***}$ & 0.117$^{**}$ & 0.290$^{***}$ & -0.117$^{**}$ \\
& (0.054) & (0.051) & (0.031) & (0.051) \\
 $\log\textsf{FeeRevenue}$ & -0.870$^{***}$ & -0.199$^{**}$ & -0.671$^{***}$ & 0.199$^{**}$ \\
& (0.131) & (0.095) & (0.086) & (0.095) \\
 \textsf{Markout} & -0.043$^{***}$ & -0.075$^{***}$ & 0.032$^{***}$ & 0.075$^{***}$ \\
& (0.012) & (0.014) & (0.010) & (0.014) \\
 \textsf{CompetitorShare} & 0.189$^{**}$ & 0.067$^{}$ & 0.122$^{***}$ & -0.067$^{}$ \\
& (0.077) & (0.069) & (0.034) & (0.069) \\
 \textsf{Internalization} & -0.000$^{}$ & 0.128$^{***}$ & -0.128$^{}$ & -0.128$^{***}$ \\
& (0.118) & (0.034) & (0.108) & (0.034) \\
\hline \\[-1.8ex]
 Observations & 39464 & 39464 & 39464 & 39464 \\
 N. of groups & 40 & 40 & 40 & 40 \\
 $R^2$ & 0.306 & 0.063 & 0.360 & 0.063 \\
\bottomrule \\[-2.25ex]
\multicolumn{5}{l}{Pool and day fixed effects are included; standard errors are clustered at the pool level.} \\
\textit{Note:} & \multicolumn{4}{r}{$^{*}$p$<$0.1; $^{**}$p$<$0.05; $^{***}$p$<$0.01}
\end{tabular}
\label{table:private-all}
\end{table}

\subsection{Large Trade Size (10 WETH)}

\vspace{-2em}

\begin{table}[H] \centering
\caption{
Baseline Regression Model ($\Delta=10$, $h_r=1$ day, $h_m=5$ mins)
}
\begin{tabular}{l 
>{\centering\arraybackslash}p{0.2\linewidth} 
>{\centering\arraybackslash}p{0.2\linewidth} 
>{\centering\arraybackslash}p{0.2\linewidth} 
>{\centering\arraybackslash}p{0.2\linewidth}}
\toprule
& (1) & (2) & (3) & (4) \\
& $\log\textsf{v3Spread}$ & $\log\textsf{cfv2Spread}$ & $\log\textsf{v3S/cfv2S}$ & $\log\textsf{TVL}$ \\
\midrule
% const & 1.383$^{***}$ & 3.016$^{***}$ & -1.633$^{***}$ & 15.257$^{***}$ \\
% & (0.000) & (0.000) & (0.000) & (0.000) \\
 $\log\textsf{GasPrice}$ & 0.319$^{**}$ & 0.158$^{}$ & 0.162$^{***}$ & -0.158$^{}$ \\
& (0.129) & (0.133) & (0.049) & (0.133) \\
 \textsf{LogReturns} & -0.040$^{***}$ & -0.010$^{}$ & -0.030$^{***}$ & 0.010$^{}$ \\
& (0.011) & (0.007) & (0.008) & (0.007) \\
 \textsf{Volatility} & 0.381$^{***}$ & 0.093$^{*}$ & 0.289$^{***}$ & -0.093$^{*}$ \\
& (0.065) & (0.052) & (0.043) & (0.052) \\
 $\log\textsf{FeeRevenue}$ & -0.950$^{***}$ & -0.232$^{***}$ & -0.717$^{***}$ & 0.232$^{***}$ \\
& (0.131) & (0.088) & (0.091) & (0.088) \\
 \textsf{Markout} & -0.091$^{***}$ & -0.183$^{***}$ & 0.093$^{***}$ & 0.183$^{***}$ \\
& (0.029) & (0.022) & (0.016) & (0.022) \\
\hline \\[-1.8ex]
 Observations & 34256 & 34256 & 34256 & 34256 \\
 N. of groups & 40 & 40 & 40 & 40 \\
 $R^2$ & 0.321 & 0.090 & 0.459 & 0.090 \\
\bottomrule \\[-2.25ex]
\multicolumn{5}{l}{Pool and day fixed effects are included; standard errors are clustered at the pool level.} \\
\textit{Note:} & \multicolumn{4}{r}{$^{*}$p$<$0.1; $^{**}$p$<$0.05; $^{***}$p$<$0.01}
\end{tabular}
\label{table:baseline-usethis}
\end{table}

\vspace{-3em}

\begin{table}[H] \centering
\caption{
Extended Regression Model ($\Delta=10$, $h_r=1$ day, $h_m=5$ mins)
}
\begin{tabular}{l 
>{\centering\arraybackslash}p{0.2\linewidth} 
>{\centering\arraybackslash}p{0.2\linewidth} 
>{\centering\arraybackslash}p{0.2\linewidth} 
>{\centering\arraybackslash}p{0.2\linewidth}}
\toprule
& (1) & (2) & (3) & (4) \\
& $\log\textsf{v3Spread}$ & $\log\textsf{cfv2Spread}$ & $\log\textsf{v3S/cfv2S}$ & $\log\textsf{TVL}$ \\
\midrule
% const & 1.383$^{***}$ & 3.016$^{***}$ & -1.633$^{***}$ & 15.257$^{***}$ \\
% & (0.000) & (0.000) & (0.000) & (0.000) \\
 $\log\textsf{GasPrice}$ & 0.270$^{**}$ & 0.148$^{}$ & 0.122$^{**}$ & -0.148$^{}$ \\
& (0.122) & (0.127) & (0.049) & (0.127) \\
 \textsf{LogReturns} & -0.039$^{***}$ & -0.010$^{}$ & -0.029$^{***}$ & 0.010$^{}$ \\
& (0.011) & (0.007) & (0.007) & (0.007) \\
 \textsf{Volatility} & 0.363$^{***}$ & 0.089$^{*}$ & 0.275$^{***}$ & -0.089$^{*}$ \\
& (0.064) & (0.052) & (0.043) & (0.052) \\
 $\log\textsf{FeeRevenue}$ & -0.881$^{***}$ & -0.203$^{**}$ & -0.678$^{***}$ & 0.203$^{**}$ \\
& (0.133) & (0.091) & (0.085) & (0.091) \\
 \textsf{Markout} & -0.093$^{***}$ & -0.184$^{***}$ & 0.090$^{***}$ & 0.184$^{***}$ \\
& (0.028) & (0.022) & (0.014) & (0.022) \\
 \textsf{CompetitorShare} & 0.220$^{***}$ & 0.051$^{}$ & 0.169$^{***}$ & -0.051$^{}$ \\
& (0.064) & (0.057) & (0.030) & (0.057) \\
 \textsf{Internalization} & 0.098$^{}$ & 0.105$^{***}$ & -0.007$^{}$ & -0.105$^{***}$ \\
& (0.094) & (0.034) & (0.072) & (0.034) \\
\hline \\[-1.8ex]
 Observations & 34256 & 34256 & 34256 & 34256 \\
 N. of groups & 40 & 40 & 40 & 40 \\
 $R^2$ & 0.335 & 0.097 & 0.475 & 0.097 \\
\bottomrule \\[-2.25ex]
\multicolumn{5}{l}{Pool and day fixed effects are included; standard errors are clustered at the pool level.} \\
\textit{Note:} & \multicolumn{4}{r}{$^{*}$p$<$0.1; $^{**}$p$<$0.05; $^{***}$p$<$0.01}
\end{tabular}
\label{table:private-all}
\end{table}

% \newpage
% \input{sections/outline}
\end{document}